\documentclass [12pt,preprint]{aastex}

\usepackage{epsfig}
\begin{document}
\voffset-1cm
\newcommand{\gsim}{\hbox{\rlap{$^>$}$_\sim$}}
\newcommand{\lsim}{\hbox{\rlap{$^<$}$_\sim$}}

\title{Hyperstars -- Main Origin of Short Gamma Ray Bursts?}
        
\author{Arnon Dar\altaffilmark{1},}

\altaffiltext{1}{arnon@physics.technion.ac.il, dar@cern.ch.\\
Physics Department and Space Research Institute, Technion, Haifa 32000,
Israel}

\begin{abstract}

The first well-localized short-duration gamma ray bursts (GRBs), GRB
050509b, GRB 050709 and GRB 050724, could have been the narrowly
beamed initial spike of a burst/hyper flare of soft gamma ray
repeaters (SGRs) in host galaxies at cosmological distances.  Such bursts
are expected if SGRs are young hyperstars, i.e. neutron stars where a
considerable fraction of their neutrons have converted to hyperons and/or
strange quark matter.

\end{abstract}

\section{Introduction}

Gamma ray bursts (GRBs) divide into two distinct classes; long-duration
($T>2\, s$) soft-spectrum bursts (long GRBs) and short-duration ($T<2\,
s$)  hard-spectrum bursts (short GRBs). There is mounting evidence from
observations of optical afterglows (AGs) of relatively nearby long
duration GRBs that they are produced by highly relativistic jets ejected
in supernova (SN) explosions of massive stars, as long advocated by the
cannonball model of GRBs (Dar \& De R\'ujula~2004 and references therein).  
However, so far, no optical AG of short GRBs has been detected and their
origin is still unknown.

The leading scenarios for the origin of short GRBs, include, (a) merger of
neutron-star (ns) or black hole (bh) binaries\footnote{Merger of 
neutron-star or
black hole binaries was first suggested as possible origins of long GRBs,
but, their localization in star-formation regions and their association
with core collapse supernovae left this scenario viable only for short
GRBs.} (Goodman et al.~1987; Eichler et al.~1989; Mochkovitch et 
al.~1993), (b)  gravitational collapse of accreting white dwarf (Dar \& De
R\'ujula~2003;2004), (c) gravitational collapse of neutron stars to
strange-quark stars (Dar~1999) or hyperstars (Dar \& De R\'ujula~2000) and
(d) giant flares from soft gamma ray repeaters (SGRs) in external galaxies
(Dado et al.~2005; Hurley et al.~2005).

So far, only SGRs are a proven source of short GRBs.  SGRs are widely
believed to be magnetars: slowly rotating neutron stars with ultra-strong
surface magnetic field, $B\sim 10^{15}\, Gauss\,,$ misaligned with respect
to their rotation axis, which spin down rapidly by magnetic dipole
radiation and whose main energy source is their magnetic field energy. In
the magnetar model of SGRs (Duncan and Thompson~1995), crustal
instabilities lead occasionally to dissipation of their magnetic energy
through large scale magnetic reconnection on their surface, which  
produce their X-ray and $\gamma$-ray flares.  However, the energy release 
in such
events cannot exceed the total magnetic field energy, $E_M\sim B^2\,
R^3/12\,,$ where $B$ is their surface magnetic field which is constrained
by their spin-down rate, and $R$ is the radius of the neutron star (ns).
As of today, about 10 magnetar candidates are known and their $B$ field, 
estimated from their spin down rate, is listed in Table 1. Even 
if the
total field energy of such magnetars was released in a single short 
duration quasi-isotropic flare, the GRB energy could not have exceeded 
$5\times 10^{46}\, erg\, .$

Recently, the SWIFT satellite (e.g. Gehrels et al.~2004)  
provided the first rapid and accurate X-ray localization of three short
GRBs, GRB 050509b, GRB 050709 and GRB 050724, which lasted 
$\lesssim 250\, {\rm ms}.$
Their fluences were quite typical of short GRBs with durations
$\lesssim 250$ ms, observed before by BATSE on board CGRO
(Paciesas et al.~1999).  Follow-up optical observations have detected a
giant, non-star-forming elliptical galaxy at redshift $z=0.225$ in the XRT
error circle of GRB 050509b  (Bloom et al.~2005), a bright star-forming 
galaxy at redshift
$z=0.16$ in the XRT error circle of GRB 050709 (Price et al.~2005)
and a bright elliptical
galaxy in the XRT error circle of GRB 050724 at $z=0.257$ 
(Berger et al.~2005), respectively,
with a small chance probability. If these short GRBs were physically
associated with these galaxies, then their isotropic equivalent energies,
estimated 
from their redshift distances (standard cosmology with
$\Omega_M\approx 0.27,$ $\Omega_\Lambda\approx 0.73$ and $h=0.65$)  and
measured fluences were (Berger et al.~2005),
$\sim 9\times 10^{49}$ erg for GRB 050509b and $\sim
4\times 10^{50}$ erg for GRB 050709 and GRB 050724. They are smaller than 
the
typical isotropic energies of long duration GRBs by $\sim$ 2-3 orders of
magnitude. 
However, these equivalent isotropic energies are larger than
both the equivalent isotropic gamma ray energy, $\sim 4\times 10^{46}$
erg, released by the giant flare from SGR 1860-20 on December 27, 2004
(Hurley et al.~2005) and the entire magnetic field energy of any 
known magnetar candidate, by more than 3 orders of magnitude (see Table
1). Because magnetars are not expected to produce highly collimated
hyper-flares, this has been considered (e.g Gehrels et al.~2005) as
conclusive evidence against the SGR origin of short GRBs at
$z > 0.01\, .$

The non detection of a supernova in deep optical images of the host galaxy
of GRB 050509b taken with large telescopes, such as Gemini 
(Bersier et al.~2005)  and VLT (Hjorth et al.~2005) up to 3 weeks after 
burst, were used
to argue (Hjorth et al.~2005) that ``the absence of an SN rules out
models \footnote{Short GRBs may have more than a single origin. 
Evidence from a single GRB may rule out certain sources
for that particular GRB, but, it cannot ``rule out'' any source for other
GRBs.} predicting a normal SNIa associated with short GRBs''.

The possible association of GRB 050509b and GRB 050724 with non-star
forming elliptical galaxies was advanced, e.g., by Gehrels et al.~(2005),
by Hjorth et al.~(2005) and by Berger et al.~(2005) as supporting evidence
for the assumption that short GRBs are produced by 
ns-ns or ns-bh mergers, despite the observation that GRB 
050709 was associated with a star-forming galaxy.

However, in this letter we argue that hyperflares of SGRs may be observed
from cosmological distances and can be the main source of the observed
short GRBs. 
The giant flares/bursts from SGR 0526-66 on March 5, 1979, SGR
1900+14 on August 27, 1998 and SGR 1806-20 on December 27, 2004 consisted
of an initial short ($<0.5\, s$) spike and a much longer ($>200\, s$)
pulsating tail. While the spikes had quite different intensities and
thermal bremsstrahlung spectra (e.g., Mazets er al.~2005; 
Palmer et al.~2005) the pulsating tails had similar durations, fluences 
and black body
spectra (e.g., Mazets et al.~1999; Hurley et al.~2005). It suggests that,
perhaps, the above three bursts/giant flares were similar bursts where
the initial spike was generated by the beamed emission of a relativistic
jet (e.g., by inverse Compton scattering of ambient light)  and the
differences were mainly because of different viewing angles of the jet,
while the similar pulsating tail was a black body surface emission with an
enhanced temperature near the polar caps. The detected radio afterglows
from the giant flares of SGR 1900+14 (Frail et al.~1999) and SGR 1806-20
(Cameron et al.~2005; Gaensler et al.~2005) provide additional evidence
that relativistic jets are ejected in giant flares of SGRs (Yamazaki et
al.~2005). In this letter we argue that SGRs are not magnetars, but are
hyperstars (Dar \& Re R\'ujula~2000), i.e. ns's where a
significant fraction of their neutrons are converting to hyperons or
strange quark matter, releasing gravitational binding energy which
accumulates and causes eruptions with ejection of collimated bipolar jets
along their magnetic axis. These jets presumably produce the initial short
and bright spike of giant flares, which appear as short GRBs from
cosmological distances. Like in ordinary pulsars, the ejection mechanism 
of the collimated jets is not clear.

\section{Hyperstars}
Ordinary nuclear matter is made entirely of neutrons and protons which
contain only valence $u$ and $d$ quarks. Baryons that contain $s$
(``strange'') quarks, such as $\Sigma$ and $\Lambda\, ,$ were first
discovered in the late 40's and early 50's of the last century and were
named `hyperons'.  It was speculated long ago that strange matter made of
$u\,,$ $d$ and $s$ quarks can be the true ground state of hadronic matter
(Bodmer~1971; Witten~1984) and that the interior of neutron-like star
consists of such deconfined quarks and not of neutrons. It was
also suggested that there may be no ns's and all neutron-like
stars are strange quark stars (Alcock et al.~1986) containing
approximately equal numbers of deconfined $u$, $d$ and $s$ quarks.
Even if strange quark matter is not the true ground state of hadronic
matter, hyper matter bound by a strong gravitational potential can still
be the ground state of nuclear matter in hyperstars -- ns's
where a significant fraction of their neutrons have converted to hyperons.
Both in ns's and in hyperstars, neutrons and hyperons do not
decay because the quantum states of their Fermionic decay products are 
already occupied in the star (Pauli blocking). Indeed,
simple arguments strongly suggest that
slowly-rotating, cold ns's have a critical mass beyond which they
collapse to hyperstars which continue to be hyperactive 
and release gravitational binding energy by gradual 
contraction, first due to the conversion $n\,n\rightarrow p\,\Sigma^-$
and later also due to $n\,n\rightarrow n\,\Lambda$ and $n\,n\rightarrow 
n\,\Sigma^0$ conversions: 

For instance, ignoring first general-relativistic corrections, the radius 
and central
density ${\rho_c}$ of a self-gravitating degenerate Fermi gas of neutrons 
of total baryonic mass M  and zero angular momentum obtained from
a (postulated) polytropic (Emden-Lane) solution of the
hydrostatic equation are,
\begin{equation}
 R\approx 15.1
       \left({M\over M_\odot}\right)^{-1/3}~km,
\label{radius}
\end{equation}
 \begin{equation}
 \rho_c\approx 6\,\bar\rho \approx 0.83\times 10^{15}
             \left({M\over M_\odot}\right)^2~g~cm^{-3}.
\label{density}
\end{equation}
In this simplest of models, low mass
ns should indeed be made of neutrons 
and a small fraction of protons and electrons to assure
stability against their $\beta$ decay, 
\begin{equation}
n_p\approx \left[{h\, c\over m_n\, c^2}\right]^3
            {3\, n_n^2\over 64\, \pi}\, .
\label{np}
\end{equation}
But as M is increased past $ \sim 1\,M_\odot$, $\rho_c$ increases until
the central Fermi energy $\epsilon_f(n)=(h^2/8\, m_n)\, (3\, \rho_c/\pi \,
m_n)^{2/3}$ exceeds $(m_{\Sigma^-} + m_p - 2\,m_n)\,c^2+ \epsilon_f(p)$.
At this point, it is favourable for the strangeness changing weak process
${\rm n\,n\to p\,\Sigma^-}$ (or ${\rm u\,d\rightarrow s\,u}$) to start
transforming neutron pairs at the top of the Fermi sea into (initially
pressureless) ${\rm \Sigma^-\,p}$ pairs at the bottom of the sea. This
reduces the pressure, causes contraction and increases ${\rho_c}$, which
initiates a run-away strangeness changing reactions which stop only when
the balance of chemical potentials of the various species, mainly
$n\,,p\,, \Lambda\,, \Sigma$ and $e^-$ guarantees $\beta$ stability.

Although this argument is based on Newtonian gravity, it is valid
also in general relativity because
general relativity produces a stronger
effective gravity at short distances, i.e. 
gravity becomes singular at
zero distance in Newtonian gravity, 
while in general relativity, it
becomes `infinite' already at the Schwartzchild radius, $R_s=2\, 
G\,M/c^2\,.$ Consequently, it yields larger central densities 
and enforces the transition to a hyper star.

In general relativity, slowly rotating ns's  satisfy the 
Tolman- Oppenheimer-Volkoff
(TOV) equations (Tolman~1939; Oppenheimer \& Volkoff~1939) for 
hydrostatic equilibrium,
\begin{equation}
{dP\over dr}= -{G\over r}{[\epsilon +P]\, [M+4\,\pi\, r^3\, P]\over 
                          [r-2\, G\, M]}\, , \label{TOV1} \end{equation}
\begin{equation} {dM\over dr}=4\, \pi\, r^2\, \epsilon\,, \label{TOV2}
\end{equation} where $P$ and $\epsilon$ are, respectively, the pressure
and total energy density in the star ($c=1$), $G$ is the gravitational
constant and $M(r)$ is the gravitational mass inside radius $r\, .$ The
TOV equations set a limit $R>(9/4)\, G\, M/c^2=1.125\, R_s$ for maximal
compact stars, which is much smaller than the radius of a canonical ns.
The actual radius of a compact star is determined by the solutions of
the TOV equations, which are rather sensitive to the equation of state
$\epsilon(\rho,\, P)$ of baryonic matter at sub-nuclear, nuclear and super
nuclear densities and pressures which are below the densities where
quantum chromodynamics (QCD) becomes asymptotically free. They cannot be
calculated yet from first principles. Thus, the precise properties of
strange quark stars and hyperstars cannot yet be predicted reliably enough
for establishing their existence from astronomical observations.  Rather
than adopting a specific model, we will assume that SGRs are hyperstars
with a gravitational mass similar to that of canonical ns's,
$M\approx 1.4\, M_\odot\,,t$ and with a radius significantly smaller than
that of ns's, and that their energy source is gravitational
contraction induced by the transition from neutron matter to hyper matter.
 
Is there supportive evidence that the slowly rotating
SGRs and anomalous X-ray pulsars (AXPs)  
are objects much more  compact than ordinary ns's? The best supportive 
evidence that the slowly
rotating SGRs and AXPs  are hyperstars with a radius significantly
smaller than that of ordinary radio pulsars, may come from the black body
component of their persistent emission (Dar and De R\'ujula~2000)  and/or
from the redshift of $e^+e^-$ annihilation line if emitted near their 
surface. The
application of the Stefan-Boltzman law to their black body emission
yields, $F_X=\sigma_{_B}\, R_\infty ^2\,T_\infty^4\, /d^2$ where
$\sigma_B$ is the Stefan-Boltzman constant, and $T_\infty $ and $R_\infty
$ are, respectively, the stellar effective surface temperature and radius
as inferred from measurements of the spectral flux density $F_X$
at a large distance $d$
from the star. The true stellar radius $R$ and the effective radius
$R_\infty$ are related through, $R=R_\infty/(1+z)$ where $(1+z)=[1-2\, G\,
M/R\,c^2]^{-1}$ is the gravitational redshift factor. Unfortunately, the
values of $R_\infty$ which may be extracted from black body fits 
to accurare spectral energy flux measurements of SGRs and AXPs
in the soft X-ray region (e.g. by  XMM) are proportional to their 
uncertain distances and are also sensitive to the extinction along the 
line of sight . 

\section{GRBs from hyperstars}
\noindent
{\bf GRBs from the birth of SGRS/hyperstars:}
The total energy release in the transition of an ns to a hyper
star is $\sim 50\%$ of the gravitational energy release, because $\sim$
half of the gravitational energy release is used to increase the pressure
and energy of the Fermi gas. The total energy release for, e.g. an
Emden-Lane polytrop is approximately  
\begin{equation} 
\Delta E\sim \left({2\, G\,M^2\over 7\, R_{hs}}\right)  
{R_{ns}-R_{hs}\over R_{ns}}\approx 5\times 10^{52}\, erg\,, 
\label{lumi} 
\end{equation} 
where the transition from an ns with a canonical ${\rm M_{ns}= 1.4\,
M_\odot}$ to a cold hyper star changes its radius from $R_{ns}\sim 10\,
km$ to $R_{hs}\sim 7\, km\, ,$ the typical radius of a hyper star. Most of
this internal release of energy can be radiated away in a very short burst
of neutrinos and antineutrinos. The dynamical time scale of the collapse
is very short, $\tau\sim 1/\sqrt{G\,\bar{\rho}}\leq ms$, yielding a
relatively high efficiency of neutrino annihilation to $e^+e^-$ pairs
outside the bare hyperstar which can produce a relativistic
$e^+,e^-,\gamma$ fireball and a GRB with an isotropic energy of
$\sim 10^{50-51}\, erg$ (Goodman et al.~1987).
Such bursts can be seen from $Gpc$ distances.
However, the birth-rate of SGRs is $\sim 1/750 \, y^{-1}$
in our Galaxy (see Table 1) and $\sim 2\times 10^6\, y^{-1}$ 
within a luminosity distance of $\sim 2\, Gpc\, ,$ 
assuming that their birth rate is proportional 
to the star formation rate ($\sim (1+z)^4\, ,$ e.g. Perez-Gonzalez et 
al.~2005 and references therein). 
Hence, if the birth of SGRs produces 
GRBs, they must be highly collimated, i.e., 
produced by highly relativistic bipolar 
jets, presumably  along their magnetic axis, like in core collapse SN
explosions. But, SGRs are young pulsars (see Table 1). As such, they may 
be born inside supernova remnants (SNR) of core collapse or accretion 
induced collapse supernovae. Thus, like 
in the cannonball model of ordinary long GRBs 
(Dar \& De R\'ujula~2004), inverse Compton scattering of light 
emitted or scattered by the SNR may produce a long GRB which 
is dimmer than ordinary long GRBs but is not associated
with an SN  akin to SN1998bw.

\noindent
{\bf Short GRBs from SGRs/hyperstars:} The left over internal thermal
energy from the birth of a hyperstar can be radiated over a long time,
similar to the cooling time of a newly born ns. Loss of angular
momentum by relativistic particle emission along open magnetic lines (Dar
and De R\'ujula~2000) and cooling reduce the centrifugal and thermal
support. The resulting contraction converts more neutrons to hyperons in
internal layers. The heat released by the gradual phase transition from
neutron matter to hyper matter may be radiated continuously or in
bursts/flares. A large energy release in the star from a phase transition
in a stellar layer (neutronization near the crust or hyperonization in
more internal layers) may result in bipolar relativistic jets along the
magnetic axis and thermal emission from the surface, which generate a
hyper flare.

A large phase transition may take place over a short time comparable to 
the dynamical time scale of
the star ($<ms$), but the duration of the burst depends both on the 
duration of the jetted ejection and on the environment 
of the hyper star. No
doubt, the bursting activity of hyperstars is a very complex phenomenon
whose theoretical study will require many more years. At present, only
rough estimates of the total energy and the spectral, temporal and angular
properties of the initial burst and the following bursting activity 
can be made. They will be described in detail elsewhere
(Dado, Dar and De R\'ujula, in preparation).

Roughly, the cooling and continuous contraction of a hyperstar
can power a
total luminosity:
\begin{equation}
L\simeq \left({2\, G\,M^2\over 7\, R}\right) {\dot R\over R}\, .
\label{lumi}
\end{equation}
For the canonical  $M= 1.4~M_\odot$ and  $R=10\, km$ a contraction 
rate of $ \dot
R \sim 2~\mu m~ y^{-1}$ (a tiny $\dot R/R\sim 2\times 10^{-9}~y^{-1}$
is  sufficient to provide the inferred total luminosity, 
$L_X \leq 10^{36}~erg~s^{-1}\, ,$ of
SGRs and anomalous X-ray pulsars (AXPs). 
But for the explicit numerical coefficient, Eq.(\ref{lumi})
should, on dimensional grounds, be approximately correct in general.

\section{Short GRBs}

If ${E_\gamma}'$ is the total energy radiated isotropically in the rest
frame of a jet moving at an angle $\theta$ relative to the line of sight
with a bulk motion Lorentz factor $\gamma=1/\sqrt{1-\beta^2}$ (Doppler
factor, $\delta=1/\gamma\, (1-\beta\, cos\theta)$), then Doppler boosting
and relativistic beaming yield a fluence,
\begin{equation}
F_\gamma\approx \delta^3\, {(1+z)\, E'_\gamma\over 4\, \pi\, D_L^2}\,,
\label{Fluence}
\end{equation}
where $z$ is the redshift and $D_L$ is the  luminosity distance 
of the SGR. The inferred `equivalent GRB isotropic
$\gamma$-ray energy' of a GRB pulse, is
\begin{equation}
E^{iso}_\gamma= {4\, \pi\, D_L^2\, F_\gamma\over 1+z}
 \approx \delta^3\, E'_\gamma\,  .
\label{Eiso}
\end{equation}
For small viewing angles, $\theta^2\ll 1,$ and large Lorentz factors,
$\gamma^2\gg 1$, 
\begin{equation}
\delta\approx {2\gamma\over 1+\gamma^2\, \theta^2}\, . 
\label{Dop}
\end{equation}
For $\gamma \gg 1$, it is by far more likely that a
beamed GRB from a Galactic SGR is observed with a viewing  angle 
$\theta \gg 1/\gamma$ than with a viewing angle $\theta \lesssim 
1/\gamma.$  For  $\theta\gg 1/\gamma$, Eqs.~(\ref{Eiso},\ref{Dop}) yield
$E^{iso}_\gamma\propto \delta^3\propto\theta^{-6}\, .$
Thus, if the initial short spike of SGR 1806-20 
on December 27, 2004 with $E^{iso}_\gamma\approx 5\times 10^{46}\, erg$
was produced by a relativistic jet, which was viewed from 
an angle $\theta\gg \gamma$,  
it could have been seen at a redshift $z=0.25\,,$ 
as a short GRB with an isotropic energy  $\sim 4\times 10^{50}\,erg\, ,$
similar to that of the short GRBs 050509b, 050709 and 050724,
if its viewing angle was smaller by a factor $\sim$ 3.  
Note that this is 
independent of the value of $\gamma$ provided that $\theta\gg 1/\gamma.$

\section{The brightness distribution of short GRBs}

Long duration GRBs seem to be produced by highly relativistic jets ejected
from mass accretion on a proto-neutron star or a black hole in core
collapse SN explosions (see, e.g. Dar and De R\'ujula~2004). They are
detected up to very large redshifts. Short duration GRBs, if produced by
SGRs are much dimmer and are detected from smaller distances where the
geometry of the Universe is nearly Euclidean. In a steady state Euclidean
universe, the number $n(>P)$ of GRBs with peak photon fluxes exceeding $P$
behaves like $P^{-3/2}\, ,$ independent of beaming, for $P\geq P_{min}$,
where $P_{min}$ is the detection threshold.
Cosmic evolution modifies this behaviour 
for large values of $z$ (low values
of $P$).  If short GRBs are relatively much nearer they should 
deviate less from a $ P^{-3/2}$ behaviour.
Figure 1. presents plots of $n(>P)$
for the long ($T_{90}>2\, s$) and short duration ($T_{90}<1\, s$) GRBs
(circles) in the 4-th BATSE catalog (Paciesas et al.~1999). A small
number (11) of GRBs with $1s<T_{90}<2s\,,$ which can belong to the tail
of either one of the two distributions, were not included
in the plots. The lines are best fitted power-laws.
The best fitted power-law indices are, $-1.45\pm 0.11$ with
$\chi^2/dof=0.33$ and $-1.42\pm 0.28$ with $\chi^2/dof=0.14$, 
respectively,
compatible with the expectation. However, the deviation from the
$n(>P)\sim P^{-3/2}$ law seen in Fig. 1 is much larger for long duration
GRBs than for short duration GRBs, consistent with our expectation.
It suggests that the distances of short GRBs in the BATSE 4-th catalog 
extend up to a much shorter distance than that of long GRBs.

\section{Conclusions}

\begin{itemize}

\item
Short GRBs from cosmological distances may be produced mainly by
hyperflares from SGRs, if the initial spike of their hyperflares is highly
beamed. Such hyperflares may be the result of phase transitions from
neutron matter to hyperon or strange-quark matter in hyperstars.

\item
The so called `peak energy ' of short GRBs ($E_p=E$ at max\{$E_\gamma^2\,
dn_\gamma/dE$\}), in the BATSE sample (Paciesas et al.~1999) is slightly
higher than that of long GRBs. However, in the rest frames of their
progenitors, they may be quite similar.  In the CB model of GRBs (e.g.,
Dar and De R\'ujula~2004), this suggests similar Lorentz factors $\gamma$
(and consequently similar beaming effects)  in short and Long GRBs. Then,
their different luminosities/equivalent isotropic $\gamma$-ray energies
may result from a smaller baryon number of the jet ejected in a hyperflare
and/or a smaller density of the ambient light (`glory') near SGRs, than
those in long GRBs which are produced in SN explosions.

\item
The shorter duration of the pulses of short GRBs relative to long GRBs
(McBreen et al.~2003) may result from a smaller scale-height of the
ambient light (glory) around SGRs. 

\item
Although anomalous pulsars (SGRs and AXPs) are
very young pulsars ($\tau\lesssim 10^4$ y), and the typical kick 
velocity of pulsars ($\sim 400\, {\rm km\, s^{-1}}$)
is much smaller than the typical velocity of the spherical ejecta in SN
explosions, only a fraction of them are found within/near young SNRs
(see Table 1). This indicate
that SGRs may be born not only in core collapse SN explosions but also in, 
e.g. accretion induced collapse of white dwarfs. This may explain why short
GRBs are produced  both in elliptical galaxies with old stellar 
populations and in star-forming spiral galaxies with young stellar 
populations.  

\item
Finally, the rate of hyperflares from Galactic SGRs 
($\sim 10^{-1}\,{\rm y^{-1}}$) is much larger than the
estimated rate of ns-ns and ns-bh  mergers in the Galaxy ($\sim
1.8\times 10^{-4}\, {\rm y^{-1}}$)  (Kalogera et al.~2004). In the CB
model, it makes SGRs a much more likely source of short GRBs than
ns-ns or ns-bh merger.
\end{itemize}

\noindent {\bf Acknowledgement:} This research was supported in 
part by the Asher Fund for Space Research at the Technion.
It is based on an ongoing collaboration with S. Dado and A. De R\'ujula
for which the author is very grateful.

\begin{table}[t]
\caption{Magnetic Field Energy of SGRs and AXPs}
\begin{tabular}{lcccccc}\tableline
Identity & $P$   & $\dot{P}^a$ & $\tau_c^b$ & $B^c$& $E_B^d$& SNR? \\      
  & (s)   &  ($10^{-11}$) &  (kyr)   & ($10^{14}$~G)&($10^{46}$~erg)& \\
\tableline
\multicolumn{5}{c}{SGRs} \\
\tableline
SGR 0525$-$66 & 8     &  6.6  &  1.9  &  7.4 & 4.5& N49 \\
SGR 1627$-$41 & 6.4?     &  ?  &   ?  &  ? &? & no      \\
SGR 1806$-$20 & 7.5   & 2.8 & 4.2 & 4.6&  1.8& no \\
SGR 1900+14   & 5.2   & 6.1 & 1.3 & 5.7 & 2.6 & no \\
\tableline
\multicolumn{5}{c}{AXPs} \\
\tableline
CXOU J011$-$7211? & 8.0 & ? & ? & ? & ?& no  \\
4U 0142+62 & 8.7 & 0.20 & 69 & 1.3& 0.14& no  \\
1E 1048.1$-$5937 & 6.4 & 3.3 & 3.0 & 4.7& 1.8& no \\
RX 1708$-$4009 & 11.0 & 1.9 & 9.2 & 4.6& 1.8& no  \\
XTE J1810$-$197? & 5.5 & 1.5 & 7.6 & 2.5 & 0.52& no  \\
1E 1841$-$0450 & 11.8 & 4.2 & 4.6 & 7.0 & 4.1& Kes 73  \\
AX 1844$-$0258? &  7.0 &  ? & ?  & ?& ? & G29.6+0.1  \\
1E 2259+586 & 7.0 & 0.048 & 220 & 0.6& 0.03& CTB 109 \\ 
\tableline 
\tableline
\end{tabular}
\\
$^a$Long-term average value.\\ 
$^b$Characteristic age estimated from $P/2\dot{P}$. \\
$^c$Surface dipolar magnetic field estimated from $3.2\times 10^{19} 
(P\dot{P})^{1/2}$~G. \\
$^d$Magnetic field energy estimated from $B^2\, R^3/12$~erg\\
 Unknown or uncertain entry denoted by  ``?'' \\
\end{table}
\newpage
\begin{figure}[t]
\hspace {1.7cm}
\epsfig{file=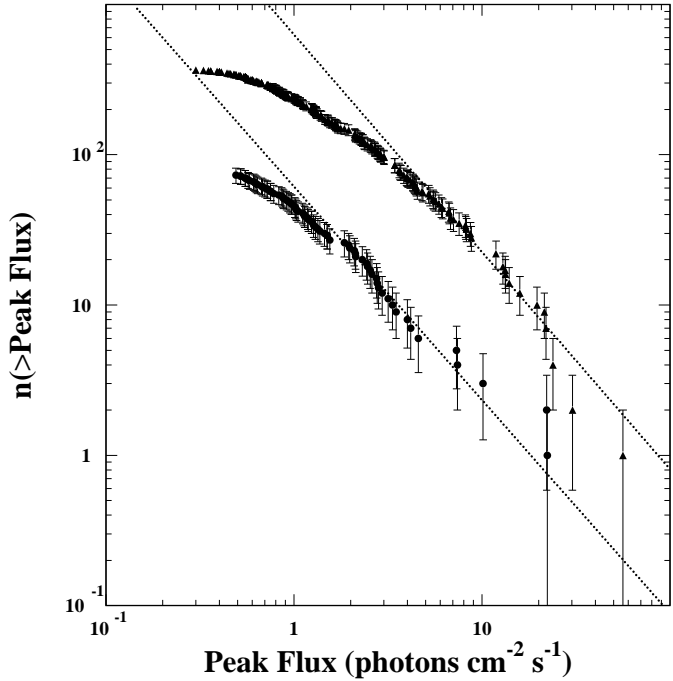, width=10cm}
\\
\caption{The number of long duration ($T_{90}>2\, s$) GRBs
(triangles) and short duration ($T_{90}<1\, s$) GRBs 
(circles) in the 4-th BATSE catalog (Paciesas et al. 1999)
with peak photon flux above the indicated value. The lines 
are the best fitted power-law to the distribution of GRBs with 
peak flux $>5\, cm^{-2}\, s^{-1} $  and  $>2\, cm^{-2}\, s^{-1}\,,$ 
respectively. 
The best fitted power-law indices are, $-1.45\pm 0.11$ with 
$\chi^2/dof=0.33$
and $-1.42\pm 0.27$ with $\chi^2/dof=0.14$, respectively. }
\label{fig1}
\end{figure}

\end{document}